\documentclass[superscriptaddress,secnumarabic,amssymb,amsmath,nobibnotes,aps,prd,showpacs,nofootinbib]{revtex4}
\usepackage{caption}
\usepackage{subcaption}
\usepackage{graphicx}
\usepackage{bm}
\usepackage{amsmath}
\usepackage{amsfonts}
\usepackage{amssymb}
\usepackage{epstopdf}
\usepackage{color}%
\providecommand{\U}[1]{\protect\rule{.1in}{.1in}}
\newcommand{\be}{\begin{equation}}
\newcommand{\ee}{\end{equation}}

\newcommand{\mincir}{\raise
-3.truept\hbox{\rlap{\hbox{$\sim$}}\raise4.truept\hbox{$<$}\ }}
\newcommand{\magcir}{\raise
-3.truept\hbox{\rlap{\hbox{$\sim$}}\raise4.truept\hbox{$>$}\ }}

\newtheorem{remark}{Remark}[section]

\begin{document}

\title{Cosmological perturbations in a class of  fully covariant modified theories: \\ 
Application to models with the same background as standard LQC}

\author{Jaume de Haro\footnote{E-mail: jaime.haro@upc.edu}}
\affiliation{Departament de Matem\`atiques, Universitat Polit\`ecnica de Catalunya, Diagonal 647, 08028 Barcelona, Spain}

\author{Llibert Arest\'e Sal\'o\footnote{E-mail: llibert.areste-salo@tum.de} }
\affiliation{Departament de Matem\`atiques, Universitat Polit\`ecnica de Catalunya, Diagonal 647, 08028 Barcelona, Spain}
\affiliation{TUM Physik-Department, Technische Universit{\"a}t M{\"u}nchen, James-Franck-Str.1, 85748 Garching, Germany}

\author{Emilio Elizalde\footnote{E-mail: elizalde@ieec.uab.es} }
\affiliation{Institut for Space Sciences, ICE/CSIC-IEEC, Campus UAB, Carrer de Can Magrans s/n, 08193 Bellaterra (Barcelona) Spain}
\affiliation{International Laboratory for Theoretical Cosmology, TUSUR University, 634050 Tomsk, Russia}

\thispagestyle{empty}

\begin{abstract}
Bouncing cosmologies are obtained by adding to the Einstein-Hilbert action a term of the form $\sqrt{-g}f(\chi)$, with $\chi$ a scalar depending on the Hubble
parameter only, not on its derivatives, and which is here shown to arise from the divergence of the 
unitary time-like eigenvector of the stress tensor. At background level, the dynamical equations for a given $f$-theory are calculated, showing that the  
simplest bouncing cosmology resulting leads to exactly the same equations as those for holonomy corrected Loop Quantum Cosmology (LQC). When dealing with perturbations, the equation for tensor ones is the same as in General Relativity (GR); for scalar perturbations, when one uses the $f$-theory which leads to the same background as 
the standard version of holonomy corrected LQC, one obtains similar equations (although a bit more elaborated) as those coming from LQC in the so-called {\it deformed algebra approach}. 

\end{abstract}

\vspace{0.5cm}


\maketitle



\section{Introduction}

One of the most simple bouncing backgrounds (see \cite{patrick0, patrick, odintsov0} for recent reviews about bounces) is obtained from holonomy corrected Loop Quantum Cosmology (LQC), where the corresponding Friedmann equation depicts an ellipse on the plane $(H, \rho)$ \cite{singh, singh1,singh2,singh3, Haro1, Haro2}, being $H$ the Hubble parameter and $\rho$ the energy density. As shown in several papers, this simple background can be mimicked by modifying the Einstein-Hilbert action through the introduction of a term of the form $\sqrt{-g}f(\chi)$ where $g$ is the determinant of the metric, $f$ is a well-known function \cite{helling, ds09, Norbert, langlois} and $\chi$ is a scalar, only depending on $H$ but not on its derivatives. The problem with this method is to actually find a scalar that for synchronous observers in the Friedmann-Lema{\^\i}tre-Robertsont-Walker (FLRW) spacetime will only depend on the Hubble parameter.

\

From our viewpoint the simplest scalar is the extrinsic curvature \cite{ha17}, which appears in a natural way when using the ADM formalism \cite{adm}. Disappointingly, this is not a covariant theory because
 the extrinsic curvature is not a true scalar in the sense that it depends on the slicing chosen and, thus, its use is only justified if there exists a preferred  foliation of the spacetime. Following the spirit of {\it Weyl's principle} (see \cite{rz} for a historical review), one could choose a preferred slicing as follows. The time-like eigenvector of the stress tensor, which always exists for realistic matter (see pages 89-90 of \cite{haw}), generates a preferred non-crossing  family of world-lines and one can construct, at any
given time $t$, a family of hypersurfaces ortogonal to these world-lines, obtaining in this way the so-called {\it co-moving slicing}.

\

Another simple scalar could be obtained by working in the Weitzenb\"ock spacetime (the usual Levi-Civita connection is replaced by the Weitzenb\"ock one) \cite{w}, where torsion does not vanish. In this spacetime, the scalar torsion for synchronous observers in the flat FLRW geometry is equal to minus three times the Hubble parameter, thus satisfying the required property. Unfortunately, as has been shown in \cite{lsb}, this theory is not locally Lorentz invariant.

 \
 
For this reason, people have kept looking for a really covariant invariant, and have dealt with the Carminati-McLenaghan invariants \cite{cm, hp18}, with a general function of the Ricci and Gauss-Bonet scalars or with second derivatives of the Riemann tensor \cite{yoshida}. The problem with these purely gravitational invariants is that they are quite involved and lead to very complicated equations for cosmological perturbations. Moreover, in our approach --not using the principle of the {\it limiting curvature hypothesis} considered in \cite{yoshida}-- they can easily lead to Ostrogradski or gradient instabilities, as well as to the appearance of ghost fields. All these problems led specialists to explore other ways, such as modified mimetic  gravity \cite{mukhanov, mukhanov1,langlois}, so as to find out such scalar. 
 
 \
  
Following these arguments, guided by Weyl's principle and taking into account that for co-moving observers in flat FLRW spacetime the divergence of a unitary time-like vector is equal to $-3H$, in the present paper we propose as our fully covariant scalar the divergence of the unitary time-like eigenvector of the stress tensor, which, in the case of a universe filled up with a scalar field $\phi$ minimally coupled to gravity, is equal to $u_{\mu}=\frac{\phi_{,\mu}}{\sqrt{\phi_{, \nu}\phi^{, \nu}}}$ (throughout the work we will use the notation: $\phi_{,\mu}\equiv\partial_{\mu}\phi=\nabla_{\mu}\phi$). 
 
\ 
 
With this covariant scalar, we will show how to construct $f$-theories leading to bouncing backgrounds and calculate the perturbed equations for scalar and tensor perturbations for a given $f$-theory. When dealing with perturbations and working in the longitudinal gauge --where the Newtonian potential, namely $\Phi$, and the variation of the scalar field, namely $\delta\phi$, are the dynamical variables-- the corresponding dynamical system turns out to be a coupled one. This is an essential difference with respect to theories such as General Relativity (GR) or LQC in the {\it deformed algebra approach}, where the dynamical equation for the potential $\Phi$ decouples ($\delta\phi$ does not appear, see for instance \cite{mukhanovbook, grain}).

\

Once these equations are obtained, we study some characteristics of the matter-ekpyrotic bounce scenario \cite{cw,haa} as the calculation of some spectral quantities and the reheating temperature via the gravitational particle production of massless particles, in the contracting regime, during the phase transition from matter domination to the ekpyrotic regime. 

\

The manuscript is organized as follows: In Section II we present our class of modified gravitational theories and obtain the corresponding dynamical equations. We study them at the background level and find which is the model that
leads to the simple bounce predicted by holonomy corrected LQC. 
A Hamiltonian analysis of our theory is performed in Section III, which leads to the conclusion that the present theory, as in the case of mimetic gravity, has one more degree of freedom than GR. 
In section IV, we study scalar and tensor perturbations. For scalar perturbations, working in the longitudinal gauge, we obtain the equations for the Newtonian potential and for the perturbed part of the scalar field, showing that they are coupled. Moreover, we derive the corresponding Mukhanov-Sasaki equations for our theory. Then, dealing with tensor perturbations, we show that, since the modification of our theory is performed on the matter sector, the equations must be the same as for GR. 
Section V is devoted to the comparison, at the perturbative level, of our model which leads to the same background as holonomy corrected LQC, with other theories which also lead to the same background, as:
LQC in the {\it deformed algebra approach} \cite{grain, caitelleau, vidotto}, teleparallel LQC \cite{cai, haro13}, extrinsic curvature LQC \cite{ha17} and mimetic LQC \cite{mukhanov1, langlois, Norbert}.
In Section VI, we study the matter-ekpyrotic scenario applied to the model that leads the same background as  LQC. We calculate the spectral index and its running and show that they match the most recent observational data. Moreover, we study the reheating process via gravitational massless particle production. Finally, the last Section is devoted to conclusions.
 
\

The units used throughout the paper are $\hbar=c=1=M_{pl}=1$, where $M_{pl}$ is the reduced Planck mass with the
convention that a temporal vector $v_{\mu}$ satisfies $v_{\mu}v^{\mu}<0$ and with the notation: 
\begin{enumerate}
\item $\varphi_{,\mu}\equiv\partial_{\mu}\varphi=\nabla_{\mu}\varphi$ for a given scalar $\varphi$.
\item  $\bar{g}$ is the unperturbed part of $g$.
\item $g_{\chi}$ means derivative of $g$  
with respect to $\chi$ and $g_{\phi}$ derivative of $g$ with respect to $\phi$, for a given function $g$.
\end{enumerate}

\section{A class of modified gravitational theories}

All models to be considered here come from a simple action, which consists in adding a term of the form $f(\chi)$ to the Einstein-Hilbert action. Here $f$ is a given function which, in order to recover GR, vanishes at low energy densities, 
while $\chi$ is a fully covariant scalar built from the scalar field that fills the whole universe and whose value in any co-moving slicing of the flat FLRW spacetime is proportional to the Hubble parameter. More precisely, we consider
\begin{eqnarray}\label{action}
 S=\int\sqrt{-g}\left(\frac{1}{2}R+f(\chi)    +{\mathcal L}_{matt}\right) d^4x,
 \end{eqnarray}
$R$ being the scalar curvature. We have assumed that the matter sector of the universe is described by a scalar field $\phi$ with a potential, $V(\phi)$ which is minimally coupled to gravity and whose Lagrangian is given by
 \begin{eqnarray}
  {\mathcal L}_{matt}= \left(-\frac{\phi_{,\mu}\phi^{,\mu}}{2}-V(\phi)\right),
  \end{eqnarray}
and
 $\chi\equiv -\nabla_{\mu} u^{\mu}$, being the vector $u^{\mu}$ the time-like unitary eigenvector of the stress tensor
 \begin{eqnarray}
 T_{\mu}^{\nu}=\phi_{,\mu}\phi^{,\nu}-\left(\frac{1}{2}\phi_{,\alpha}\phi^{,\alpha}+V(\phi)   \right)\delta_{\mu}^{\nu}, \end{eqnarray}
 that is, 
 \begin{eqnarray}
 u^{\mu}=\frac{\phi^{,\mu}}{\sqrt{-\phi^{,\alpha}\phi_{,\alpha}}}.
 \end{eqnarray}




The dynamical equations are obtained by performing the variation of the action with respect to $g_{\mu\nu}$, leading to  
\begin{eqnarray}\label{dyn}
G_{\mu\nu}=T_{\mu\nu}+\tilde{T}_{\mu\nu},
\end{eqnarray}
where $G_{\mu\nu}=R_{\mu\nu}-\frac{1}{2}g_{\mu\nu}R$ is the well-known Einstein tensor, while the tensor $\tilde{T}_{\mu,\nu}$, coming from the term $f(\chi)$, is given by
\begin{eqnarray}
\tilde{T}_{\mu\nu}
\equiv
\left(f-\chi f_{\chi}+u^{\alpha}\chi_{,\alpha}f_{\chi\chi} 
\right)g_{\mu\nu}-f_{\chi\chi}(u_{\nu}\chi_{,\mu}+u_{\mu}\chi_{,\nu}
+u^{\alpha}\chi_{,\alpha}u_{\mu}u_{\nu}).
\end{eqnarray}


\


On the other hand, the variation of the action with respect to the scalar field $\phi$ leads to the following conservation equation,
\begin{eqnarray}\label{conser}
-\Box \phi+ V_{\phi}-\nabla_{\mu}\left(  \frac{1}{\sqrt{-\phi^{,\alpha}\phi_{,\alpha} }} \left( \partial^{\mu}f_{\chi}+u^{\mu}u_{\alpha}\partial^{\alpha}f_{\chi}   \right) \right)=0,
\end{eqnarray}
which differs from the usual one  $-\Box \phi+ V_{\phi}=0$. However, for the FLRW geometry it leads to the standard conservation equation,
\begin{eqnarray}\label{conserb}
\ddot{\phi}+3H\dot{\phi}+V_{\phi}=0.
\end{eqnarray}

\


\

\subsection{The background}

Considering backgrounds (i.e., solutions of (\ref{conserb})) satisfying $\dot{\bar{\phi}}(t)>0$ all the time, 
the simplest way to obtain the dynamical equations is as follows: We consider the flat metric $ds^2=-N^2(t)dt^2+a^2(t)\delta_{ij}dx^idx^j$, 
which leads to $\chi=\frac{3H}{N}$, where $N(t)$ is the lapse function. Hence,
after integration by parts, the action becomes
\begin{eqnarray}
S={\mathcal V}\int a^3N\left(-\frac{3H^2}{N^2}+ \bar{f}\left(\frac{3H}{N}  \right)+{\mathcal L}_{matt}         \right) dt,
\end{eqnarray}
where ${\mathcal V}$ is the volume of the fixed  elementary spatial cell, where all spatial integrations are performed,  and  where the matter Lagrangian simplifies to
\begin{eqnarray}\label{Lagrangian}
{\mathcal L}_{matt} =\frac{{\dot{\bar{\phi}}}^2}{2N^2}-V(\bar\phi).
\end{eqnarray}

Performing the variation with respect to $N$ and taking at the end $N=1$, one obtains the modified Friedmann equation for synchronous observers
\begin{eqnarray}\label{modfriedman}
\rho=3H^2+\bar f -3H\bar f_{\chi},
\end{eqnarray}
which depicts a curve on the plane $(H,\rho)$. Finally, taking its temporal derivative and using the conservation  equation (\ref{conserb}) in the form $\dot{\rho}=-3H(P+\rho)$, one finds  the Raychaudhuri equation
\begin{eqnarray}\label{raychauduri}
\left(1-  \frac{3}{2}\bar f_{\chi\chi} \right)\dot{H}=-\frac{1}{2}(\rho+P).
\end{eqnarray}

\
\begin{remark}
A different  way to obtain such equations is to directly consider the metric for synchronous observers $ds^2=-dt^2+a^2\delta_{ij}dx^idx^j$ and use the equations $0-0$ and $i-i$, which also leads to the modified Friedmann and Raychaudhuri equations.
\end{remark}

\

Therefore, given a curve $\rho=\bar g(3H)=\bar g(\bar\chi)$ on the plane $(H,\rho)$, so as to obtain the corresponding $\bar f(\bar\chi)$ theory one has to solve the first-order differential equation
\begin{eqnarray}
\bar\chi \bar f_{\chi}-\bar f-\frac{1}{3}\bar\chi^2+g(\bar\chi)=0,
\end{eqnarray}
whose solution is
\begin{eqnarray}\label{formula}
\bar f(\bar\chi)=+\frac{1}{3}\bar\chi^2-\bar\chi\int\frac{g(\bar\chi)}{\bar\chi^2}d\bar\chi.
\end{eqnarray}

\

Note that, in order to obtain a bouncing background,
a necessary condition is to choose a curve on the plane $(H,\rho)$ cutting two or more times the axis $H=0$. One of those points has to be  $(0,0)$
because  at low energy densities a viable bouncing background has to approach to GR, i.e., the chosen curve has to approach to the parabola $H^2=\frac{\rho}{3}$ and the other
points cutting the axis $H=0$ have the form $(0, \rho_i)$ with $\rho_i>0$.

 The simplest example
is the ellipse coming from 
the holonomy corrected Friedmann equation in standard  LQC \cite{singh06,svv06,singh08} (Note that here we are not dealing with the recent model of LQC proposed by
 Dapor and Liegener in \cite{dl})
\begin{eqnarray}\label{flqc}
\rho=\bar g(\bar\chi)=\frac{\rho_c}{2}\left(1\pm\sqrt{1-\frac{4\bar\chi^2}{3\rho_c}}   \right)\Longleftrightarrow H^2=\frac{\rho}{3}\left(1-\frac{\rho}{\rho_c}\right),
\end{eqnarray}
where the so-called critical density has the value $\rho_c\sim 0.4 \rho_{pl}=0.4\times 64\pi^2\cong 252$ (see for instance \cite{singh3}).   From this equation one can 
see that the Hubble parameter vanishes at $\rho=0$ and $\rho=\rho_c$, i.e., the  points cutting the axis $H=0$ are  $(0,0)$ and $(0,\rho_c)$, as one can see from Figure $1$.

\begin{figure}[h]
\begin{center}
\includegraphics[scale=0.5]{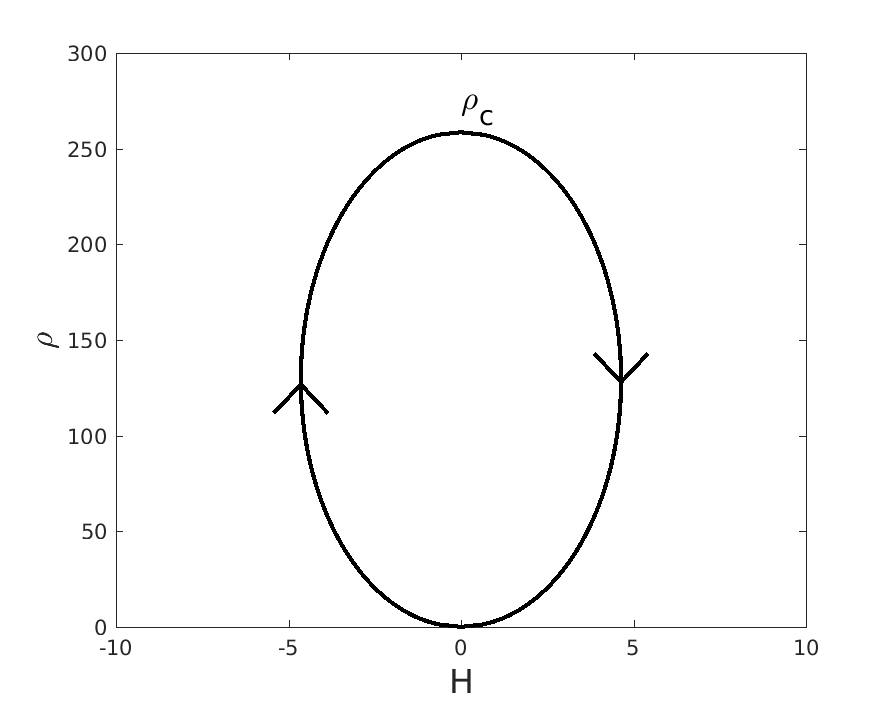}
\end{center}
\caption{Ellipse depicted by the Friedmann equation  in standard LQC, and its dynamics for a non-phantom  fluid with linear Equation of State (EoS) $P=w\rho$.}
\end{figure}



\

For this background, using equation (\ref{formula}), the corresponding $f$-theory is given by
\begin{eqnarray}\label{fLQC}
\bar f(\bar \chi)=\frac{1}{3}\bar\chi^2+\frac{\rho_c}{2}\left(1-\sqrt{1-s^2}-s\arcsin s  \right),
\end{eqnarray}
where $s\equiv \frac{2}{\sqrt{3\rho_c}}\bar\chi$ and the functions $\sqrt{1-s^2}$  and  $\arcsin s$ are bi-valued  \cite{hp18}. For the theory to be well defined, 
since the ellipse has two branches -the upper part corresponding to 
$\rho=\frac{\rho_c}{2}\left(1+\sqrt{1-\frac{4\bar\chi^2}{3\rho_c}} \right)$ and the lower one
$\rho=\frac{\rho_c}{2}\left(1-\sqrt{1-\frac{4\bar\chi^2}{3\rho_c}} \right)$-, we have to choose a convenient prescription. For example, we can choose
the sign of the square root as positive (respectively negative) in the lower (respectively upper) branch and
$ \arcsin s\equiv \int_0^s \frac{1}{\sqrt{1-{\bar s}^2}}  {d\bar{s}}$ in the lower branch, whereas  $\arcsin s\equiv \int_0^s \frac{1}{\sqrt{1-{\bar s}^2}}  {d\bar{s}}+\pi$, in the upper one, with the same criteria for the sign of the square root, thus obtaining that the function $\bar f$ is continuous in all the ellipse.

\

Note also that, a simple calculation shows that, for this particular $f$-theory,  the modified Friedmann and Raychaudhuri equations read
\begin{eqnarray}\label{LQCstandard}
H^2=\frac{\rho}{3}\left(1-\frac{\rho}{\rho_c}\right) \quad \mbox{and}\quad \dot{H}=-\frac{\rho+P}{2}\left(1-\frac{2\rho}{\rho_c}\right),
\end{eqnarray}
and coincide with the ones obtained  in holonomy corrected LQC \cite{singh09,sst06}. 

\

A final remark is in order. Since in our case the matter sector is depicted by a non-phantom scalar field $\bar{\phi}$ (see equation (\ref{Lagrangian})), 
we will have infinitely many backgrounds because the conservation equation 
$\ddot{\bar{\phi}}+3H\dot{\bar\phi}+V_{\phi}(\bar\phi)=0$
is a second order differential equation. Effectively, for any initial condition $\bar{\phi}(0)=\alpha_0$ and  $\dot{\bar{\phi}}(0)=\alpha_1$, one has a different background. 
Therefore, if we choose a curve $\rho=g(\bar\chi)$ in the plane $(H,\rho)$
containing a point of the form $(0, \tilde{\rho})$ with $\tilde{\rho}>0$ and initial conditions satisfying $ \frac{\alpha_1^2}{2}+V(\alpha_0)=\tilde{\rho}$,
its corresponding solution will lead to a bouncing background provided that $V_{\phi}(\alpha_0)\not= 0$.

\

As an example one can consider the matter bounce scenario in LQC, which is given by the potential 
$V(\phi)=2\rho_c\frac{e^{-\sqrt{3}\phi}}{(1+e^{-\sqrt{3}\phi})^2}$  \cite{ha14a}.
In this case, the conservation equation has the following analytic solution
\begin{equation}\label{solution}
\bar{\phi}=\frac{2}{\sqrt{3}}\ln\left( \sqrt{\frac{3}{4}\rho_c}t+\sqrt{\frac{3}{4}\rho_ct^2+1}  \right),
\end{equation}
leading to the following background
\begin{eqnarray}\label{background}
a(t)=\left(\frac{3}{4}\rho_ct^2+1  \right)^{1/3}, \quad H=\frac{\frac{1}{2}\rho_ct}{\frac{3}{4}\rho_ct^2+1 }, \quad \rho=\frac{\rho_c}{\frac{3}{4}\rho_ct^2+1 },
\end{eqnarray}
which is the same nonsigular bouncing background obtained solving the equation (\ref{LQCstandard}) when one considers a pressureless fluid and whose  dynamics is depicted in Figure $1$. All the other solutions, which lead to different
backgrounds, 
are obtained choosing different initial conditions (see Figure $3$ of \cite{ha17}, where it is showed that nearly all solutions lead to nonsingular bouncing backgrounds).

\

Moreover, since we are dealing with theories beyond GR, in order to have a bouncing background it is not needed to violate the {\it null energy condition} $\rho+P\geq 0$ near the bounce
using quintom or Lee-Wick matter (see \cite{easson} and references therein) because the bounce occurs when the value of the energy density is strictly positive.
For example, looking at (\ref{background}) one can see that at the bouncing time $t=0$ the energy density is given by $\rho_c$ and the pressure, which could be obtained
from the Raychaudhuri equation (\ref{LQCstandard}), is zero. So, the null energy condition is fulfilled at the bounce.

\section{Hamiltonian analysis}

In this section we perform a Hamiltonian analysis in order to find the degrees of fredoom of our model. To this end, we will use
 the ADM formalism \cite{adm}, where the line element acquires the form 
\begin{eqnarray}
 ds^2=g_{\mu\nu}dx^{\mu}dx^{\nu}= -N^2dt^2+\gamma_{ij}(dx^i+N^idt)(dx^j+N^jdt),
 \end{eqnarray}
the matter Lagrangian becomes  
  \begin{eqnarray}
  {\mathcal L}_{matt}
  =\left(\frac{\dot{\phi}^2}{2N^2}-\frac{N^i}{N^2}\dot{\phi}\phi_{,i}-\frac{1}{2}\left(\gamma^{ij}-\frac{N^iN^j}{N^2}\right)\phi_{,i}\phi_{,j}-V(\phi)\right),
  \end{eqnarray}
and the action is given by 
\begin{eqnarray}
 S=\int_{-\infty}^{\infty}\left\{\int_{\Sigma_t} N\sqrt{\gamma} \left(\frac{1}{2}({\mathcal R}+{\mathcal I})
 +f(-\nabla_{\mu}u^{\mu})+{\mathcal L}_{matt}  \right)d^3x\right\} dt , 
 \end{eqnarray}
where
 ${\mathcal R}$ is the intrinsic curvature, i.e. the scalar curvature of $\Sigma_t$ , and ${\mathcal I}= K_{ij}K^{ij}-(K_i^i)^2$ is the extrinsic curvature scalar,
 being 
 \begin{eqnarray}
 K_{ij}=\frac{1}{2N}\left(D_iN_j+D_jN_i-\dot{\gamma}_{ij}     \right)
 \end{eqnarray}
 the extrinsic curvature tensor, 
 with $D$  the induced Levi-Civita connection in the slicing $\Sigma_t$.
 
 \
 
Introducing a Lagrangian multiplier field, namely $\beta$, the action becomes
\begin{eqnarray}\label{action1}
 S=\int_{-\infty}^{\infty}\left\{\int_{\Sigma_t} N\sqrt{\gamma} \left(\frac{1}{2}({\mathcal R}+{\mathcal I})
 +f(\chi)+ \beta(\chi+ \nabla_{\mu}u^{\mu} )+ {\mathcal L}_{matt}  \right)d^3x\right\} dt=\nonumber\\ 
 \int_{-\infty}^{\infty}\left\{\int_{\Sigma_t} N\sqrt{\gamma} \left(\frac{1}{2}({\mathcal R}+{\mathcal I})
 +f(\chi)+ \beta\chi-  \beta_{, \mu}u^{\mu} + {\mathcal L}_{matt}  \right)d^3x\right\} dt, 
 \end{eqnarray}
with the canonical momenta being
\begin{eqnarray}
P^{ij}=\frac{1}{2}\sqrt{\gamma}(K^{ij}-K^l_l\gamma^{ij}), \quad P_{\beta}=\frac{\sqrt{\gamma}}{N}(u_0-N^iu_i), \quad 
P_{\phi}=\frac{\sqrt{\gamma}}{N}(\dot{\phi}-\phi_{,i}N^i)-\frac{N\sqrt{\gamma}}{\sqrt{-\phi_{,\nu}\phi^{,\nu}}}(\beta^{,0}+u^0u^{\mu}\beta_{,\mu}),
\end{eqnarray}
 and the constraints 
\begin{eqnarray}
P_{\chi}\approx 0, \quad P_{N}\approx 0, \quad P_{N^i}\approx 0. 
\end{eqnarray}

Note that, from the equation $P_{\beta}=\frac{\sqrt{\gamma}}{N}(u_0-N^iu_i)$, we obtain $\dot{\phi}$ as a function of $P_{\beta}$. In fact, taking the square of this expression
one gets $-\phi_{,\nu}\phi^{,\nu}=\frac{\gamma \gamma^{ij}\phi_{,i}\phi_{,j}}{{P^2_{\beta}-\gamma}}$ and, thus, 
\begin{eqnarray}
\dot{\phi}=\frac{N\sqrt{\gamma^{ij}\phi_{,i}\phi_{,j}}P_{\beta}}{\sqrt{P_{\beta}^2-\gamma}}+N^i\phi_{,i}.
\end{eqnarray}
And, from $P_{\phi}=\frac{\sqrt{\gamma}}{N}(\dot{\phi}-\phi_{,i}N^i)-\frac{N\sqrt{\gamma}}{\sqrt{-\phi_{,\nu}\phi^{,\nu}}}(\beta^{,0}+u^0u^{\mu}\beta_{,\mu})$,
we readily obtain $\dot{\beta}$ as a function of $P_{\beta}$ and $P_{\phi}$.

What is important is that, after the Legendre transformation, one can get the Lagrangian as follows,
\begin{eqnarray}
H=\int_{\Sigma_t} (N{\mathcal H}+N^i{\mathcal H}_i+{\mathcal H}_{matt}+\alpha_NP_N+\alpha_{N^i}P_{N^i}+\alpha_{\chi}P_{\chi})d^3x,
\end{eqnarray}
where ${\mathcal H}_{matt}$ is the matter part of the Hamiltonian;  $\alpha_N$, $\alpha_{N^i}$ and $\alpha_{\chi}$ are Lagrange multipliers, and
the functions ${\mathcal H}$ and ${\mathcal H}_i$  lead to the hamiltonian and diffeomorphism 
constraints, which come from imposing stability under time evolution of the constraints $P_{N}\approx 0$ and $P_{N^i}\approx0$, i.e., 
$\dot{P}_N=\{P_N,H\}={\mathcal H}\approx 0$ and $\dot{P}_{N^i}=\{P_{N^i},H\}={\mathcal H}_i\approx 0$. 

\

Now we examine the stability of the constraint $P_{\chi}$. Looking for the only term in the Hamiltonian where $\chi$ appears and for the term where it appears in the action  (\ref{action1}), i.e.
 $-N\sqrt{\gamma}(f(\chi)+\beta\chi)$, one has $\dot{P}_{\chi}=\{P_{\chi},H\}=-N\sqrt{\gamma}(f'(\chi)+\beta)$, which leads to the constraint
$C_{\chi}\equiv f'(\chi)+\beta \approx 0$. Finally, since $\dot{C}_{\chi}=\{C_{\chi},H\}=N\sqrt{\gamma}f''(\chi)\alpha_{\chi}$, the stability of $C_{\chi}$ is ensured by fixing the Lagrange multiplier $\alpha_{\chi}$ as $\alpha_{\chi}=0$.

\

Summing up, we have obtained the constraints ${\mathcal H}\cong 0$, ${\mathcal H}_i\cong 0$, $P_{\chi}\cong 0$ and $C_{\chi}\cong 0$, and the canonical pairs $(q_{ij}, P^{ij})$, $(\phi, P_{\phi})$, $(\chi, P_{\chi})$ and $(\beta, P_{\beta})$. Then, from the constraints $P_{\chi}\cong 0$ and $C_{\chi}\cong 0$ one may remove two variables, for example $P_{\chi}\cong 0$ and $\beta\approx -f'(\chi)$, thus obtaining, as in mimetic gravity \cite{Norbert}, one more degree of freedom than in the case of GR. However, as we will show in next Section, when dealing with perturbations in longitudinal gauge, the degrees of freedom are the Newtonian potential $\Phi$  and the perturbation of the scalar field $\delta\phi$ for scalar perturbations and two degrees for the tensor ones, exactly the same as in GR. This is the same as what happens in mimetic gravity \cite{hap}, where the degrees of freedom are the perturbed part of the mimetic field and $\delta\phi$.

\section{Perturbations}

In this section we will calculate, for a given $f$-theory,  the scalar and tensor perturbations using for scalar perturbations the longitudinal gauge (see for a review of the used setup \cite{mfb, cgk}).

\subsection{Scalar perturbations}

In Newtonian gauge the the line element is given by \cite{mukhanovbook}
\begin{eqnarray}
ds^2=-(1+2\Phi)dt^2+(1-2\Psi)a^2\delta_{ij}dx^idx^j.
\end{eqnarray}
where the potentials $\Phi$ and $\Psi$ coincide with the gauge invariant ones.

\

A simple calculation leads to  $\delta u_0=\Phi$ and $\delta u_k=\frac{\partial_k \delta\phi}{\dot{\bar{\phi}}}$, where $\delta\phi$, in this gauge,  coincides with the
$\delta\phi^{gi}$ (the gauge invariant perturbation of the scalar field). Then, at linear order, we have 
\begin{eqnarray}
\chi= 3H-3(\dot{\Psi}+H\Phi)-\frac{1}{a^2\dot{\bar{\phi}}}\Delta\delta\phi.
\end{eqnarray}

{

\

To obtain the dynamical equations for perturbations, note first of all that 
the perturbed  $i-j$ equation  of (\ref{dyn}) for $i\not= j$ leads to the identity $\Psi=\Phi$. Thus, 
perturbing  the
$i-0$, $i-i$ and $0-0$ equations
of (\ref{dyn}) one gets, respectively,  \begin{eqnarray}
\dot{\Phi}+H\Phi=\frac{1}{2}\dot{\bar\phi}\delta\phi-\frac{\bar f_{\chi\chi}}{2}\delta\chi,
\end{eqnarray}
\begin{eqnarray}
2\left(  \ddot{\Phi}+4H\dot{\Phi}+(3H^2+2\dot{H})\Phi
\right)=-\dot{\bar\phi}(\Phi\dot{\bar\phi}-\delta\dot{\phi})-\bar V_{\phi}\delta\phi +\delta(f-\chi f_{\chi}+u^{\alpha}\chi_{,\alpha}f_{\chi\chi}),
\end{eqnarray}
\begin{eqnarray}
2\left(3{ H}^2\Phi+3{ H}\dot{\Phi}-\frac{1}{a^2}\Delta \Phi\right)=\dot{\bar\phi}(\Phi\dot{\bar\phi}-\delta\dot{\phi})-\bar V_{\phi} \delta\phi
+\delta(f-\chi f_{\chi}-u^{\alpha}\chi_{,\alpha}f_{\chi\chi})+\Phi\dot{\bar f}_{\chi}-\partial_t(\bar{f}_{\chi\chi}\delta\chi).
\end{eqnarray}
Adding equations $0-0$ and $i-i$ and using $i-0$, one gets
\begin{eqnarray}
\ddot{\Phi}-\frac{1}{a^2}\Delta\Phi+H\dot{\Phi}+2\dot{H}\Phi = \ddot{\bar\phi}\delta\phi+
\frac{1}{2}\Phi\dot{\bar f}_{\chi}-\frac{1}{2}\partial_t(\bar{f}_{\chi\chi}\delta\chi),
\end{eqnarray}
where we have introduced the notation  $\Omega\equiv \frac{1}{1-\frac{3}{2}\bar{f}_{\chi\chi}}$.

We may write equation $i-0$ as:
\begin{eqnarray}
\ddot{\bar\phi}\delta\phi=\frac{2\ddot{\bar\phi}}{\dot{\bar\phi}}\frac{\dot{\Phi}+H\Phi}{\Omega}-\frac{\ddot{\bar\phi}}{\dot{\bar\phi}}\bar f_{\chi\chi}\frac{\Delta\delta\phi}{a^2\dot{\bar\phi}},
\end{eqnarray}
which leads to the following equation for the potential $\Phi$,
\begin{eqnarray} \label{lqcperturbed}
{\ddot{\Phi}}-\frac{\Omega}{a^2}\Delta\Phi+\left(H-2\frac{\ddot{\bar\phi}}{\dot{\bar\phi}}-\frac{\dot{\Omega}}{\Omega}\right){\dot{\Phi}}+\left(2\left(\dot{H}-H\frac{\ddot{\bar\phi}}{\dot{\bar\phi}}\right)-H\frac{\dot{\Omega}}{\Omega} \right){\Phi}=
\frac{\Omega{\dot{\bar{\phi}}}^2}{2}\partial_t\left(\frac{\bar{f}_{\chi\chi}\Delta\delta\phi}{a^2{\dot{\bar{\phi}}}^3}   \right),
\end{eqnarray}
which, for the case of the choice of $f$ given in (\ref{fLQC}), differs from the corresponding  equation of LQC in the {\it deformed algebra approach} \cite{grain, caitelleau} only in the right hand side term, which in the last approach vanishes.

\

On the other hand, the equation for $\delta\phi$ is obtained from the linearization of the conservation equation (\ref{conser})
\begin{eqnarray}\label{linearization}
\delta\ddot{\phi}+3H\delta\dot{\phi}-\frac{1}{a^2}\Delta \delta\phi+V_{\phi\phi}\delta\phi-4\dot{\bar{\phi}}\dot{\Phi}+2V_{\phi}\Phi=
\frac{\Omega\dot{\bar{\phi}}}{a^2}\Delta\left( \frac{\bar{f}_{\chi\chi}\Delta\delta\phi}{a^2{\dot{\bar{\phi}}}^3} \right).
\end{eqnarray}
}
From these equations,  we will calculate the Mukhanov-Sasaki (M-S) equation for scalar perturbations in our approach. First of all, note that equation $i-0$ can be written as
\begin{eqnarray}
\frac{d}{dt}\left(\frac{a\Phi}{H}  \right)=\frac{a\Omega{\dot{\bar{\phi}}}^2}{2H^2}
\left[\frac{H\delta\phi}{\dot{\bar{\phi}}}+\Phi+\frac{H}{a^2{\dot{\bar{\phi}}}^3}\bar{f}_{\chi\chi}\Delta\delta\phi  \right].
\end{eqnarray} 
 On the other hand, equation $0-0$ takes the form
 \begin{eqnarray}
2\left(3{ H}^2\Phi+3{ H}\dot{\Phi}-\frac{1}{a^2}\Delta \Phi\right)=\dot{\bar\phi}(\Phi\dot{\bar\phi}-\delta\dot{\phi})-\bar V_{\phi} \delta\phi-
3H\bar{f}_{\chi\chi}\delta\chi,
\end{eqnarray}
and, after a somewhat cumbersome calculation,
one can see that it is equivalent to the following one,
\begin{eqnarray}
\frac{1}{a^2}\Delta \Phi=\frac{{\dot{\bar{\phi}}}^2}{2H}\frac{d}{dt}\left[\frac{H\delta\phi}{\dot{\bar{\phi}}}+\Phi+
\frac{H}{a^2{\dot{\bar{\phi}}}^3}\bar{f}_{\chi\chi}\Delta\delta\phi  \right]-\frac{{\dot{\bar{\phi}}}^2}{2}\frac{d}{dt}\left( \frac{\bar{f}_{\chi\chi}\Delta\delta\phi}{a^2{\dot{\bar{\phi}}}^3 } \right).
\end{eqnarray}

Introducing gauge-invariant  variables (recall we are working in the Newtonian gauge) and using the conformal time
\begin{eqnarray}
v=a\left(\delta\phi+\frac{{\bar{\phi}'}}{{\mathcal H}}\Phi +  \frac{1}{{{(\bar{\phi}')}}^2}\bar{f}_{\chi\chi}\Delta\delta\phi  \right), \quad z=a\frac{{\bar{\phi}'}}{{\mathcal H}}, \quad
u=\frac{2a\Phi}{{\bar{\phi}'}\sqrt{\Omega}}, \quad \mbox{and} \quad \theta=\frac{1}{z\sqrt{\Omega}},
\end{eqnarray} 
where  ${\mathcal R}\equiv \frac{v}{z}=\Phi-\frac{H}{\dot{H}}(\dot{\Phi}+H\Phi)$ is the {\it curvature fluctuation in co-moving coordinates}, 
one obtains the following M-S equations
 \begin{eqnarray}
 \sqrt{\Omega}\Delta u=z\left(\frac{v}{z}\right)'-{{{\bar{\phi}}'}}\left( \frac{a\bar{f}_{\chi\chi}\Delta\delta\phi}{{(\bar{\phi}')}^3 } \right)' 
 ,\qquad \theta\left(\frac{u}{\theta}\right)'=\sqrt{\Omega} v,
 \end{eqnarray}
 which, after inserting the second into the first one, lead to
the equation for the potential (\ref{lqcperturbed}) in the simple form 
\begin{eqnarray}
u''-\Omega\Delta u-\frac{\theta''}{\theta}u=\bar{\phi}'\sqrt{\Omega}\left( \frac{a\bar{f}_{\chi\chi}\Delta\delta\phi}{({\bar{\phi}')}^3 }  \right)',
\end{eqnarray}
while the equation for the variable $v$, after taking the Laplacian from the second equation and using the first one, becomes
\begin{eqnarray}\label{v}
v''-\Omega\Delta v-\frac{z''}{z}v=\frac{1}{z}\left(z\bar{\phi}'\left( \frac{a\bar{f}_{\chi\chi}\Delta\delta\phi}{({\bar{\phi}')}^3 }  \right)'\right)'.
\end{eqnarray}

\

Note that in our approach,  the right hand side of Eq.~(\ref{v}) does not vanish, meaning that the variable $v$, which encodes the scalar perturbations (it depends on $\Phi$
and $\delta\phi$), is not independent and one needs another equation in order to calculate the evolution of the scalar perturbations. Fortunately in the matter (or matter-ekpyrotic) scenario, in the contracting phase, 
the pivot scale leaves the Hubble radius at rather low energy densities, as compared to the Planck one, so the corrections due to $f$ can be safely disregarded and, thus, $v$ satisfies approximately the usual equation $v''-\Delta v-\frac{z''}{z}v=0$. This finally means that
the calculation of the spectral quantities, such as the spectral index, its running, and the ratio of tensor to scalar perturbations, can safely be done using GR in the contracting phase, as we will show in next section. 

\

On the other hand, to calculate the evolution of the scalar perturbations through time, we need two equations, since the equations for $\Phi$ and $\delta\phi$ are coupled. In this case, solving the system of Eqs.~(\ref{lqcperturbed}) and (\ref{linearization}), we will obtain such evolution. The best suited variables for that are $u$ and $\delta\sigma\equiv a\delta\phi$, with the corresponding dynamical equations  being
\begin{eqnarray} \label{Z}
\left\{\begin{array}{ccc}
u''-\Omega\Delta u-\frac{\theta''}{\theta}u & = &\bar{\phi}'\sqrt{\Omega}\left( \frac{\bar{f}_{\chi\chi}\Delta\delta\sigma}{({\bar{\phi}')}^3 }  \right)' , \\
 & & \\
\delta\sigma''-\Delta \delta\sigma+a^2V_{\phi\phi}\delta\sigma-2{\bar{\phi}'}a\left(\frac{{\mathcal H}u}{a^2\theta} \right)'+\frac{a{\mathcal H}V_{\phi}u}{\theta} &=&
{\Omega{\bar{\phi}'}}\Delta\left( \frac{\bar{f}_{\chi\chi}\Delta\delta\sigma}{{({\bar{\phi}}')}^3} \right). \end{array}\right.
\end{eqnarray}
We now have to look for the initial conditions for a matter or matter-ekpyrotic bouncing scenario \citep{we, cw,haa}. Using that at very early times we are in the framework of GR, in Fourier space, we will have 
$v_k\rightarrow \frac{e^{-ik\tau}}{\sqrt{2k}}$ when $\tau\rightarrow -\infty$. At very early times, equation $\Delta u=z\left(\frac{v}{z}\right)'$
holds and, since we are in a matter domination epoch, we have $u_k\rightarrow i\frac{e^{-ik\tau}}{k\sqrt{2k}}$. Finally, from the relation 
\begin{eqnarray}
v=\delta\sigma + 
\frac{({\bar\phi}')^2}{2{\mathcal H}}
u\cong 
\delta\sigma+
\frac{3}{2}{\mathcal H}u\cong\delta\sigma,
\end{eqnarray}
one obtains the well-known result $\delta\sigma_k \rightarrow \frac{e^{-ik\tau}}{\sqrt{2k}}$ when $\tau\rightarrow -\infty$. Thus, the asymptotic conditions at very early times
are $u_k= i\frac{e^{-ik\tau}}{k\sqrt{2k}}$ and $\delta\sigma_k=\frac{e^{-ik\tau}}{\sqrt{2k}}$.

\

A final remark is in order. In a bouncing scenario, as for instance the matter-ekpyrotic one, at early times GR holds, meaning that $\bar{f}_{\chi\chi}$ could be disregarded and thus obtaining for the variable $v$, in Fourier space, the classical
equation $v_k''+(k^2-\frac{a''}{a})v_k=0$. Then, when the pivot scale leaves the Hubble radius, i.e. in the long wavelength approximation ($k^2\ll a^2 H^2$), one can safely disregard the Laplacian terms which appear on the right hand side of Eq.~(\ref{v}) (see the end of section $4$ in \cite{ha14a}).
And maybe the same happens with the right hand side of our Eq.~(\ref{v}) because it contains a Laplacian. If so, Eq.~(\ref{v}), in the long wavelength   approximation, will become, as usual, $v_k''-\frac{z''}{z}v_k=0$ and, for the $f$-theory given by (\ref{fLQC}), we will recover
the results obtained in LQC using the {\it deformed algebra approach} \cite{we,cw, haa}. Anyway, this has to be properly checked by solving numerically the system of Eqs.~(\ref{Z}).


\

However, we want to stress that the system (\ref{Z}) can in fact be solved iteratively, taking the right hand side term as a perturbation (when the pivot scale is well inside the Hubble radius, which  happens at very early times, the right hand side term can be dismissed, since GR holds, and after the pivot scale
leaves the Hubble radius, i.e. in the long wavelength approximation, the Laplacian terms  can be dismissed too). In fact, at very early times, disregarding the right hand term,  since the universe is matter-dominated, the equation for $u$ is, in Fourier space,
\begin{eqnarray}\label{short}
u''_k+k^2u_k-\frac{6}{\tau^2}u_k=0.
\end{eqnarray}
Its solution satisfying the asymptotic condition $u_k= i\frac{e^{-ik\tau}}{k\sqrt{2k}}$ reads
\begin{eqnarray}\label{z1}
u^{(0)}_k=\frac{1}{k}\sqrt{\frac{-\pi\tau}{4}}H^{(1)}_{\frac{5}{2}}(-k\tau)=\frac{i e^{-ik\tau}}{k\sqrt{2k}}\left(1-\frac{3}{k^2\tau^2}-\frac{3i}{k\tau}   \right),
\end{eqnarray}
where $H_{\frac{5}{2}}^{(1)}$ is a Hankel's function, and the super-index $(0)$ means zero order approximation. On the other hand, when the pivot scale has left the Hubble radius, i.e. when $k^2\tau^2\ll 1$, one has 
\begin{equation}\label{long}
u''_k-\frac{\theta''}{\theta}u_k=0.
\end{equation}

\

Note that, 
in the case of the matter bounce scenario, given by the potential
$V=2\rho_c\frac{e^{-\sqrt{3}\phi}}{(1+e^{-\sqrt{3}\phi})^2}$, and choosing
the solution (\ref{solution}) which leads to the background (\ref{background}) mimicking the same background as a pressureless fluid,
one can see that the function $\frac{\theta''}{\theta}$, which is symmetric with respect to the contracting and expanding phase, satisfies that in the contracting regime is increasing from $\rho=0$, where it vanishes, to $\rho=\frac{\rho_c}{2}$, and decreasing 
from $\rho=\frac{\rho_c}{2}$ to the bounce, which occurs at $\rho=\rho_c$, and where  one has $\left|\frac{\theta''}{\theta}\right|=\frac{5}{4} \rho_c\cong 315$,
which, as we will see in  Section VI, is bigger than the pivot scale $k_*\sim 5\times 10^{-33}a_E\ll \rho_c$. Then, the pivot scale, which occurs at a very early time, namely $t=-t_H$, reenters to the Hubble radius at time $t=t_H$, so in this case the equation (\ref{long}) holds from $-t_H$ to $t_H$ and the equation (\ref{short}) holds at very early and late times.

\

The  solution of (\ref{long}) is, then,
\begin{eqnarray}\label{z2}
u^{(0)}_k=C_1(k)\theta+C_2(k)\theta\int_{\tau_I}^{\tau}\frac{1}{\theta^2}d\bar\tau  \Longrightarrow \Phi_k^{(0)}=\frac{H}{2a\theta}u_k^{(0)}
=C_1(k)\frac{{\mathcal H}}{2a^2}+C_2(k)\left(1-\frac{\mathcal H}{\mathcal H_I}\left(\frac{a_I}{a} \right)^2-\frac{{\mathcal H}\int_{\tau_I}^{\tau} a^2d\bar\tau}{a^2}\right), 
\end{eqnarray}
 where the sub-index $I$ refers to an early time where GR holds, and the coefficients $C_1(k)$ and $C_2(k)$ are obtained by matching both expressions of $u_k$ in (\ref{z1}) and 
(\ref{z2}) when $k^2\tau^2\ll 1$.

\

Once this solution has been obtained, we consider the $i-0$ equation,  which can be written as follows,
\begin{eqnarray}\label{z3}
\frac{1}{\Omega}({\Phi}'_k+{\mathcal H}\Phi_k)=\frac{\bar{\phi}'}{2a}\left(1-\frac{k^2}{ (\bar{\phi}')^2  }\bar{f}_{\chi\chi}\right)\delta\sigma_k.
\end{eqnarray}
Inserting $\Phi_k^{(0)}$ on it, one gets $\delta\sigma_k^{(0)}$. Then, to find the new iteration of $u_k^{(1)}$, one has to solve the equation 
\begin{eqnarray}
u_k''+k^2 u_k-\frac{\theta''}{\theta}u_k & = &-k^2\bar{\phi}'\sqrt{\Omega}\left( \frac{\bar{f}_{\chi\chi}\delta\sigma_k^{(0)}}{({\bar{\phi}')}^3 }  \right)',
\end{eqnarray}
using the well-known {\it method of variation of constants} for  second-order differential equations (see for example  Chapts. $13-17$ of \cite{levinson}). When one has $u_k^{(1)}$ one calculates $\Phi_k^{(1)}$ and,
inserting the result in (\ref{z3}), one gets $\delta\sigma_k^{(1)}$. And, thus, the new iterations are iterativelly obtained and a fully fledged method is constructed.

\

Alternatively, one could also deal directly with Eq.~(\ref{v}), considering the right hand side as a perturbation. In Fourier space, at very early times this equation becomes
\begin{eqnarray}
v''_k+k^2v_k-\frac{2}{\tau^2}v_k=0,
\end{eqnarray}
whose solution satisfying the corresponding asymptotic is
\begin{eqnarray}\label{xxx}
v_k^{(0)}=\frac{e^{-ik\tau}}{\sqrt{2k}}\left(1-\frac{i}{k\tau}\right).
\end{eqnarray} 
When the pivot scale leaves the Hubble radius one can use the longwavelenght approximation (for the matter bounce scenario using the background (\ref{background}) the function 
 $\frac{z''}{z}$ is symmetric with respect to the expanding and contracting phase, being an increasing function in the whole contracting period. Thus, if the pivot scale leaves the Hubble radius at $t=-t_H$ it reenters at $t=t_H$)
\begin{equation}
v''_k-\frac{z''}{z}v_k=0,
\end{equation}
whose solution is \cite{haro13}
\begin{eqnarray}
v^{(0)}_k=B_1(k)z+B_2(k)z\int_{-\infty}^{\tau}\frac{1}{z^2}d\bar\tau.
\end{eqnarray}
At early times, if one deals with the matter bounce scenario with the background  (\ref{background}),  this expression is equal to $v_k^{(0)}= -\frac{B_1(k)}{4\sqrt{3}}\rho_c\tau^2+\frac{4B_2(k)}{\sqrt{3}\rho_c}\frac{1}{\tau}$, which one has to match with
(\ref{xxx}) so as to obtain
\begin{eqnarray}
B_1(k)=\sqrt{\frac{8}{3}}\frac{ k^{3/2}}{\rho_c} \quad \mbox{and} \quad B_2(k)=-i\sqrt{\frac{3}{8}}\frac{\rho_c}{2 k^{3/2}}.
\end{eqnarray}
When one has the expression of $v_k^{(0)}$ one can use its definition to write $\Phi_k^{(0)}$ as a function of $\delta\sigma_k^{(0)}$. Then, inserting this expression into Eq.~$i-0$, one obtains a solvable first-order differential equation in $\delta\sigma_k^{(0)}$. To obtain the next iteration one has to solve, using the method of variation of constants for second order differential equations, the equation 
  \begin{eqnarray}
v_k''+ \left(k^2-\frac{z''}{z}\right)v_k=-\frac{k^2}{z}\left(z\bar{\phi}'\left( \frac{\bar{f}_{\chi\chi}\delta\sigma_k^{(0)}}{({\bar{\phi}')}^3 }  \right)'\right)'.
\end{eqnarray}
Once $v_k^{(1)}$ has been obtained, one has  to use the definition of $v$ and Eq.~$i-0$ to find $\delta\sigma_k^{(1)}$, and so continue successively to obtain the next iteration.
 

\

For example, we take  the background (\ref{background}). Since 
 from the equation (\ref{z3}) one can see that the  function $\delta\sigma_k^{(0)}(t)$ has two simple poles, one in the expanding  and the other in the contracting phase, namely $t_+>0$ and $t_-=-t_+$, to regularize the integral
one has to add an imaginary part to the poles obtaining $t_++i\epsilon$ and $t_--i\epsilon$, then after an integration by parts one will have 
\begin{eqnarray}\label{zzz}
v_k^{(1)}=B_1(k)z+B_2(k)z\int_{-\infty}^{\tau}\frac{1}{z^2}d\bar\tau
-k^2z \int_{-\infty}^{\tau}\frac{\bar{\phi}'}{z}\left( \frac{\bar{f}_{\chi\chi}\delta\sigma_k^{(0)}}{({\bar{\phi}')}^3 }  \right)' d\bar{\tau}\cong 
B_2(k)z\int_{-\infty}^{\infty}\frac{1}{z^2}d\bar\tau,
\end{eqnarray}
 due, as we will show at the end of Section VI, to the small value of the pivot scale. 
One can also see it taking into account that
\begin{eqnarray}
B_1(k)\sim \frac{k^{3/2}}{\rho_c}, \quad B_2(k)\int_{-\infty}^{\infty}\frac{1}{z^2}d\bar\tau\sim \frac{\sqrt{\rho_c}}{k^{3/2}}, \quad
k^2 \int_{-\infty}^{\tau}\frac{\bar{\phi}'}{z}\left( \frac{\bar{f}_{\chi\chi}\delta\sigma_k^{(0)}}{({\bar{\phi}')}^3 }  \right)' d\bar{\tau}\sim \frac{k^{7/2}}{\rho_c^2}.
\end{eqnarray} 
 Then, since the pivot scale $k_*$ leaves the Hubble radius at time $-t_*$,  when holonomy corrections could be disregarded, that is, when $-t_*\gg \frac{1}{\sqrt{\rho_c}}$, which
 is equivalent to $k_*=a(t_*)H(t_*)\ll \sqrt{\rho_c}$, one can see that the dominant term in (\ref{zzz}) is $B_2(k)z\int_{-\infty}^{\infty}\frac{1}{z^2}d\bar\tau$.

Therefore, we have obtained approximately the same value as in the first iteration, which coincides with the result  of LQC \cite{we, haro13}, meaning that in this iteration one will have 
\begin{eqnarray}
{\mathcal P}\equiv \frac{k^3}{2\pi^2}\left|\frac{v_k}{z}  \right|^2\cong\frac{\rho_c}{576}.
\end{eqnarray}


\subsection{Tensor perturbations}

The metric for tensor perturbations is \cite{mukhanovbook}
\begin{eqnarray}
ds^2=-dt^2+a^2(\delta_{ij}-h_{ij})dx^idx^j,
\end{eqnarray}
where $h_{ij}$ is a symmetric,  traceless and transverse tensor ($h_i^i=\partial_ih^{ij}=0$).
 


\


Since the field $\phi$ does not affect the tensor perturbations, it is clear that in this theory this equation will coincide with the corresponding one in GR, that is,
\cite{mukhanovbook}
\begin{eqnarray}
\ddot{h}_i^j+3H\dot{h}_i^j-\frac{1}{a^2}\Delta {h}_i^j=0.
\end{eqnarray}
Denoting $h_i^j$ by $h$ and introducing the variable $v_T=a h$,  in Fourier space this equation becomes 
\begin{eqnarray}
v_{T, k}''+\left(k^2-\frac{a''}{a}  \right)v_{T,k}=0.
\end{eqnarray}

\section{Comparison with other models}

We will now compare our approach with a number of other different models, all of them sharing as a background the same as for holonomy corrected LQC. These models are: LQC in the {\it deformed
algebra approach}, the teleparallel LQC approach, the intrinsic curvature LQC approach and the mimetic LQC approach. First, we shall review the dynamical equations for each of these theories:
\begin{enumerate}
\item LQC in the {\it deformed algebra approach} \cite{grain, caitelleau}.

In this case  the equation for the potential $\Phi$ decouples. It is given by 
\begin{eqnarray} 
{\ddot{\Phi}}-\frac{\Omega}{a^2}\Delta\Phi+\left(H-2\frac{\ddot{\bar\phi}}{\dot{\bar\phi}}-\frac{\dot{\Omega}}{\Omega}\right){\dot{\Phi}}+\left(2\left(\dot{H}-H\frac{\ddot{\bar\phi}}{\dot{\bar\phi}}\right)-H\frac{\dot{\Omega}}{\Omega} \right){\Phi}=0.
\end{eqnarray}
The variable M-S variable $v$ adquires the usual form 
$v=a(\delta\phi+\frac{{\bar{\phi}'}}{{\mathcal H}}\Phi) $
and the  M-S equations decouple in the simple form
\begin{eqnarray}
u''-\Omega\Delta u-\frac{\theta''}{\theta}u=0, \quad v''-\Omega\Delta v-\frac{z''}{z}v=0.
\end{eqnarray}
In particular the equation  $v''-\Omega\Delta v-\frac{z''}{z}v=0$ can be solved when the pivot scale is well inside and outside of the Hubble radius, obtaining
the whole evolution of the variable $v$, which encodes all the information about scalar perturbations and, thus, the knowledge of the power spectrum for the {\it curvature fluctuation in co-moving coordinates}.

On the other hand, for tensor perturbation the corresponding M-S equation is \cite{vidotto}
\begin{eqnarray}
h''-\Omega\Delta h-\frac{z_T''}{z_T}h=0,
\end{eqnarray}
where $z_T\equiv \frac{a}{\sqrt{\Omega}}$.

\item Teleparallel LQC \cite{haro13, cai}.

In teleparallel LQC the equation for the Newtonian potential also decouples
\begin{eqnarray}\label{pot1}
{\ddot{\Phi}}-\frac{c_s^2}{a^2}\Delta\Phi+\left(H-2\frac{\ddot{\bar\phi}}{\dot{\bar\phi}}-\frac{\dot{\Omega}}{\Omega}\right){\dot{\Phi}}+\left(2\left(\dot{H}-H\frac{\ddot{\bar\phi}}{\dot{\bar\phi}}\right)-H\frac{\dot{\Omega}}{\Omega} \right){\Phi}=0,
\end{eqnarray}
where the square of the velocity of sound is  $c_s^2={\Omega}\sqrt{\frac{\rho_c}{12 H^2}}\arcsin\left(\sqrt{\frac{12H^2}{\rho_c}} \right)$, which, contrary to what occurs in  LQC 
in the {\it deformed algebra approach}, is always positive, meaning that in this approach there are no gradient inestabilities. In the same way, the M-S also decouple and  the only difference with the ones of LQC in the {\it deformed algebra approach} is that  the velocity of sound is the same that appears in (\ref{pot1}). As in LQC in the {\it deformed algebra approach}, the equation for the variable $v$ decouples, which allows us to know the power spectrum of the {\it curvature fluctuations in co-moving coordinates}.

\

For tensor perturbations the velocity of sound is equal to $1$,   and the M-S equation is given by 
\begin{eqnarray}
h''-\Delta h-\frac{z_T''}{z_T}h=0,
\end{eqnarray}
where $z_T\equiv a  \sqrt{ \sqrt{\frac{\rho_c}{12 H^2}}\arcsin\left(\sqrt{\frac{12H^2}{\rho_c}} \right)} $.

\item Intrinsic curvature LQC \cite{ha17}.

In this approach the equations for scalar perturbations are the same as in LQC in the {\it deformed algebra approach}. However, for tensor perturbations, the corresponding
M-S equation is 
\begin{eqnarray}
h''-c_T^2\Delta h-\frac{z_T''}{z_T}h=0,
\end{eqnarray}
where $z_T$ is the same as in teleparallel LQC, and in this case the square of the velocity of sound is given by
$c_T^2= \frac{\sqrt{\frac{12H^2}{\rho_c}}}{\arcsin\left(\sqrt{\frac{12H^2}{\rho_c}} \right)}$. 

\item Mimetic LQC \cite{hap}.

This is also a fully covariant approach where, at the perturbative level, the dynamical variables are the perturbed mimetic field, namely $\delta\varphi$, and the perturbed
scalar field $\delta\phi$. In the Newtonian gauge, the potential is related with the mimetic field, as follows $\Phi=\delta\dot{\varphi}$, and the dynamical equation, as in our approach,  becomes coupled \cite{hap}
\begin{eqnarray}\label{system2}
 \left\{\begin{array}{ccc}
 \delta\ddot{\varphi}+H\delta\dot{\varphi}-\frac{c_s^2}{a^2}\Delta\delta\varphi &=&\frac{\Omega}{2}\dot{\bar\phi}\delta\phi\\
 \delta\ddot{\phi}+3H\delta\dot{\phi}-\frac{1}{a^2}\Delta \delta\phi+(V_{\phi\phi}-2{\dot{\bar\phi}}^2{{}\Omega})\delta\phi &=& 
 \frac{4{{}\dot{\bar\phi}}c_s^2}{a^2}\Delta\delta\varphi
 -2(2{{}H\dot{\bar\phi}}+V_{\phi})\delta\dot{\varphi},
 \end{array}\right.
 \end{eqnarray}
 where the square of the velocity of sound is given, as in \cite{fgm17}, by 
 \begin{eqnarray}
 c_s^2=\frac{\Omega}{2}\bar{f}_{\chi\chi}=\frac{\frac{1}{2}\bar{f}_{\chi\chi}}{1-\frac{3}{2}\bar{f}_{\chi\chi}},
 \end{eqnarray}
 which  exhibits the well-known gradient instability of the mimetic gravity case \cite{hnk17, fgm17} (see also \cite{odintsov} for the study of perturbations in specific mimetic  matter models).
 
 \
 
 Dealing with tensor perturbations, since the mimetic field does not alter the gravitational sector, as in our approach, the equations for tensor perturbations are the same as in GR.

\end{enumerate}

\

Consequently, 
we can see that for  non-fully covariant theories, such as LQC in the {\it deformed algebra approach}, teleparallel LQC, or intrinsic curvature LQC, the  equations for scalar 
perturbations decouple, which actually simplifies the theory a lot, allowing us to calculate the corresponding power spectrum. On the contrary, for the fully covariant theories, i.e. our approach and mimetic LQC, the equations of scalar perturbations do not decople, which makes their analytic study more difficult and only a numerical analysis seems to be viable in order to understand their evolution.  However, the clear advantage of the covariant theories is that the equation for tensor perturbations is the simplest one because it coincides with the one for tensor perturbations in GR.

\section{Reheating and the calculation of the spectral parameters in the bouncing matter-ekpyrotic scenario}

In this section we consider the background given by holonomy corrected LQC. In other words, we consider the function $f$ given by  (\ref{fLQC}), which means that the universe bounces when its energy density is $\rho_c$. We will show how to calculate the reheating temperature of the universe via gravitational particle production due to a phase transition from the matter domination to an ekpyrotic era, and how  the theoretical values of the spectral index and its running match well with the corresponding observational data.

\

As we will immediately show,  in a viable bouncing scenario  the pivot scale leaves the Hubble radius in the contracting phase, when GR does hold. Correspondingly,  the spectral index and its running are given,  respectively, by \cite{eho, lw}
\begin{eqnarray}
n_s= 1+12w_*,\qquad \alpha_s=\frac{12w'_*{\mathcal H}_*}{{\mathcal H}'_*},
\end{eqnarray}
where $w$ is the effective Equation of State (EoS) parameter and the ``star" means that the quantities are evaluated when the pivot scale leaves the Hubble radius. 

\

First of all, note that in order to match the theoretical value of the spectral index with the observational data, $w_*$ has to be negative and close to zero. Now, assuming that, at very early times, we have a quasi-matter domination epoch ($\dot{\phi}^2\cong 2V\rightarrow \ddot{\phi}\cong V_{\phi}$), the background equations become
\begin{eqnarray}
\left\{\begin{array}{ccc}
{\mathcal H}^2 &=& \frac{2}{3}a^2V\\
3{\mathcal H}\phi'+2a^2V_{\phi} &=& 0.
\end{array}\right.
\end{eqnarray}
In this case 
\begin{eqnarray}
w\equiv \frac{P}{\rho}=-\frac{2}{3}\left(\frac{1}{2}+\frac{{\mathcal H}'}{{\mathcal H}^2}  \right)\cong \frac{1}{3}\left(\frac{V_{\phi}}{V}  \right)^2-1
\end{eqnarray} and, thus,
\begin{eqnarray} n_s=4\left(\frac{V_{\phi_*}}{V_*}  \right)^2-11,
\qquad \alpha_s=48\left(\frac{V_{\phi_*}}{V_*}  \right)_{\phi_*}.
\end{eqnarray}  
Note that for a potential corresponding exactly to matter domination, i.e. for $V=\lambda e^{\sqrt{3}\phi}$, one obtains an exactly flat spectrum ($n_s=1$ and $\alpha_s=0$).   

We consider here a phase transition, in the contracting phase, from matter domination to an ekpyrotic regime. 
To this end, we choose a model with potential given by
\begin{eqnarray}\label{pot}
V(\phi)=\left\{\begin{array}{ccc}
\lambda e^{\sqrt{3}\phi}\left(1-\frac{2e^{\phi/k}}{2-e^{\phi/k}}\right)& \mbox{for}& \phi<0 \\
\bar{\lambda}\frac{e^{3\phi}}{(1+\frac{\bar{\lambda}}{2\rho_c}e^{3\phi})^2}& \mbox{for}& \phi\geq 0,
\end{array}\right.
\end{eqnarray}
where $k$ is a positive parameter and $\lambda$ and $\bar{\lambda}$ satisfy 
\begin{eqnarray}
-\lambda=\frac{\bar{\lambda}}{(1+\frac{\bar{\lambda}}{2\rho_c})^2} \Longrightarrow \bar{\lambda}\cong -\lambda,
\end{eqnarray}   
since, at the end of the phase transition,  we assume $|\bar{\lambda}|\ll \rho_c \cong 252$.     
This models depicts for $\phi\rightarrow -\infty$ a  matter dominated universe, and for $\phi\geq 0$ an ekpyrotic universe with
 EoS parameter $w=2$ \cite{wilson0}.

  \
  
Then, for $\phi$ negative satisfying $e^{\phi/k}\ll 1$, one has
\begin{eqnarray}
n_s\cong 1-\frac{8\sqrt{3}}{k} e^{\phi_*/k}, \qquad \alpha_s =-\frac{48}{k^2} e^{\phi_*/k}.
\end{eqnarray}
Using now the BICEP2/Keck Array and Planck  observational data at $1\sigma$ C.L., one has $n_s=0.968\pm 0.006$ and $\alpha_s=-0.003\pm 0.007$  \cite{planck}. Then, 
choosing for instance $k=10$, one can see that for $-42.38\leq \phi_*\leq -24.49$ the theoretical values of the spectral index and its running
belong to the 1-dimensional marginalized $2\sigma$ C.L..


   
  \subsection{Reheating} 
Here we will consider a reheating process due to the gravitational particle production of massless particles minimally coupled with gravity
during a phase transition from a matter dominated regime to an ekpyrotic one in the contracting phase. 
Recall that 
in our model (\ref{pot}) we have an ekpyrotic phase with EoS parameter $w=2$. Let $H_E$ be the value of the Hubble parameter at the beginning of this phase. Then, from the relations $V=3H^2+\dot{H}$ and $\dot{H}=-\frac{9}{2}H^2$, we obtain $\lambda= \frac{3}{2}H_E^2$. On the other hand, in holonomy corrected LQC a viable value of $H_E$ is approximately $-10^{-3}$ \cite{cw,haa}, which justifies our choice $|\bar{\lambda}|\ll \rho_c$.

 \
 
 The energy density of the produced particles during this phase transition is given by \citep{pv}
 \begin{eqnarray}\label{energydensity}
 \rho_r(a)=RH_E^4\left(\frac{a_E}{a} \right)^4,
 \end{eqnarray}
 where $a_E$ is the value of the scale factor at the beginning of the ekpyrotic phase and $R\sim 10^{-2}N_s$, being $N_s$ the number of different scalar fields.

\

\begin{remark}
The equation (\ref{energydensity}) was obtained considering a phase transition, in the expanding phase,  from the de Sitter phase to another one with constant EoS parameter $w>1/3$ \cite{ford,DV}. In the case we consider a phase transition in the contracting phase from the matter domination phase to another one with constant EoS parameter
$w>1$,  this formula is also valid due to the duality, pointed out in \cite{wands}, between the de Sitter regime in the expanding phase and the matter domination in the contracting one.
\end{remark}

\   
   
On the other hand, as has been showed numerically in \cite{cw, haa} (see  the figures  $11$, $13$ and $15$ of \cite{haa}), in the matter-ekpyrotic bounce scenario, after the bounce the universe enters in a kination regime, i.e., the effective EoS parameter is equal to $1$. Then, to clarify ideas we consider the background equations corresponding to LQC (\ref{LQCstandard}) and we consider a fluid with the following EoS,
\begin{eqnarray}\label{eos}
P(\rho)=\left\{\begin{array}{ccc}
0 & \mbox{for} & 0\leq \rho\leq \rho_E\equiv 3H_E^2\\
2\rho & \mbox{for} & \rho_E< \rho\leq \rho_c,
\end{array}\right.
\end{eqnarray} 
in the contracting phase, and $P(\rho)=\rho$ in the expanding one.
   
   Since for the EoS $P=w\rho$ in LQC the Hubble parameter evolves as $H(t)=\frac{\frac{(1+w)}{2}\rho_c(t-\bar{t})}{\frac{3}{4}(1+w)^2(t-\bar{t})^2+1}$ \cite{haa}, for our EoS (\ref{eos}) we will have 
 \begin{eqnarray}\label{background2}
 H(t)=\left\{\begin{array}{ccc}
 \frac{\frac{1}{2}\rho_c(t-\bar{t})}{\frac{3}{4}(t-\bar{t})^2+1}& \mbox{for} & t\leq t_E\\
 \frac{\frac{3}{2}\rho_c t}{\frac{27}{4}t^2+1}& \mbox{for} & t_E\leq t\leq 0\\
 \frac{\rho_c t}{3t^2+1}& \mbox{for} &  t\geq 0,
 \end{array}\right.
\end{eqnarray}   
 where $t_E$ is the phase transition time and, thus, it has to satisfy  $\frac{\frac{3}{2}\rho_c t_E}{\frac{27}{4}t_E^2+1}=H_E$, and $\bar{t}$ has to be chosen imposing continuity at $t=t_E$, that is, it has to satisfy  $\frac{\frac{1}{2}\rho_c(t_E-\bar{t})}{\frac{3}{4}(t_E-\bar{t})^2+1}=H_E$.

\  
   
 \begin{remark}
 Since in LQC when considering a fluid with EoS $P=w\rho$ the reconstruction method (see Section $3$ of \cite{haa} for a detailed discussion) leads to the potential 
 \begin{eqnarray}
 V(\phi)=2\rho_c(1-w)\frac{e^{\sqrt{3(1+w)}\phi}}{(1+e^{\sqrt{3(1+w)}\phi}   )^2},
 \end{eqnarray}
then the conservation equation with the following potential
 \begin{eqnarray}
 V(\phi)=\left\{ \begin{array}{ccc}
 2\rho_c\frac{e^{\sqrt{3}\phi}}{(1+e^{\sqrt{3}\phi}   )^2}& \mbox{for} & \phi\leq \phi_E\\
 -2\rho_c\frac{e^{{3}\phi}}{(1+e^{{3}\phi}   )^2}& \mbox{for} & \phi_E<\phi\leq \phi_B\\
 0 & \mbox{for} & \phi> \phi_B,
 \end{array}\right.
 \end{eqnarray}
where $\phi_E$ and $\phi_B$ satisfy respectively $2V(\phi_E)=\rho_E$ and $-2V(\phi_B)=\rho_c$,  has a solution which leads to   the background  (\ref{background2}).
 \end{remark}  
 
   \

Now, since for a fluid with the linear Equation of State $P=w\rho$ the conservation equation  $d(\rho a^3)=-Pd(a^3)$ leads to
\begin{eqnarray}
\frac{d\rho}{\rho}=-3(1+w)\frac{da}{a}\Longrightarrow \rho=\rho_i\left(\frac{a_i}{a}\right)^{3(1+w)},
\end{eqnarray}  
during the ekpyrotic phase the energy density of the background evolves as $\rho_b(a)=3H_E^2\left(\frac{a_E}{a} \right)^9$,
and, after the bounce, since the universe enters in a kination phase, the  background evolves as $\rho_b(a)=\rho_c\left(\frac{a_c}{a} \right)^6$, 
where $a_c=\left(\frac{3H_E^2}{\rho_c}  \right)^{\frac{1}{9}}a_E$, which means that after the bounce one has 
$\rho_b(a)=\rho_c\left(\frac{3H_E^2}{\rho_c}  \right)^{\frac{2}{3}}\left(\frac{a_E}{a} \right)^6$. Then, the universe
will become reheated when both energy densities are of the same order, i.e. $\rho_r(a_{reh})\sim \rho_b(a_{reh})$. 
Since in the contracting phase the energy density of the background increases faster than the one of the produced particles, the reheating will occur in the expanding phase,  when 
$\left(\frac{a_E}{a_{reh}} \right)^2\sim \frac{R H_E^4}{\rho_c\left(\frac{3H_E^2}{\rho_c}  \right)^{\frac{2}{3}}}$, and it will be
\begin{eqnarray}
T_{reh}\sim \left(\rho_b(a_{reh})\right)^{1/4}=R^{3/4}\frac{|H_E|^3}{\sqrt{\rho_c}\left(\frac{3H_E^2}{\rho_c}  \right)^{\frac{1}{3}}}.
\end{eqnarray}
Then, for $N_s\sim 1$ (GUT theories), $\rho_c\cong 252$ and $H_E\sim -10^{-3}$ one obtains $T_{reh}\sim 8\times 10^{-10}$. Finally, since $M_{pl}\cong 2.4\times 10^{18}$ GeV,  in natural units one has  $T_{reh}\sim 2\times 10^{9}$ GeV.

\begin{remark}
It is possible to obtain a lower reheating temperature by increasing the EoS parameter in the ekpyrotic phase. For example, if in the ekpyrotic phase one takes $w=5$, then the reheating temperature is reduced by one order.
\end{remark}

\

Once we have calculated the reheating temperature we can show that the pivot scale, in the contracting phase, leaves the Hubble scale when GR holds. The pivot scale is related with its physical value by $k_*=a_0 k_{phys}(t_0)$, where the sub-index $0$ means present time, and we choose, as usual, $k_{phys}(t_0)\sim 10^2 H_0\sim 10^{-59}$.

On the other hand, as we have showed at the reheating time, i.e., when both energy densities are of the same order, one has $a_{reh}\sim \sqrt{\frac{\rho_c\left(\frac{3H_E^2}{\rho_c}  \right)^{\frac{2}{3}}}{R H_E^4}}a_E\sim 5\times 10^4 a_E$. Now, from the conservation of the entropy we have the adiabatic relation  $a_0\sim \frac{T_{reh}}{T_0}a_{reh}$ \cite{rg}
and using that the current and reheating temperature are respectively $T_0\sim 8\times 10^{-32}$ and $T_{reh}\sim 8\times 10^{-10}$, one gets
$a_0\sim 5\times 10^{26} a_E$. As a consequence, $k_*\sim 5\times 10^{-33}a_E$, which means, since $|H_E|\sim 10^{-3}$, that $k_*\ll |H_E|a_E$. In other words, the pivot scale
leaves in the contracting phase the Hubble radius well after the phase transition, more precisely when $H_*\sim 5\times 10^{-33}\frac{a_E}{a_*}\leq 5\times 10^{-33}$ (in the contracting phase $a_E<a_*$) and, thus, since $\rho_c\cong 252$, one can safely disregard the effects of the $f$ theory when the pivot scale leaves the Hubble radius.

\section{conclusions}
We have constructed a class of modified gravitational theories based on the addition to the Einstein-Hilbert action of a function $f$, which depends on the divergence of the unitary time-like eigenvector of the stress tensor. We have obtained in this way a fully covariant theory, which, as in the case of mimetic gravity,  has one more degree of freedom than GR. The main advantage, at the background level, of our class of models is that, for the 
FLRW geometry, this divergence is minus three times the Hubble parameter, which allows, by choosing the function $f$ appropriately, to obtain very simple bouncing backgrounds, as the one obtained in holonomy corrected LQC.

\

At the level of cosmological perturbations, working in the Newtonian gauge, the equations for scalar perturbations exhibit some of the same interesting features as those appearing in LQC in the so-called {\it deformed algebra approach}. However,  they are not exactly the same, owing to the fact that our theory is fully covariant, in contrast with LQC in the {\it deformed algebra approach}
\cite{bojowald1,bojowald2,achour}. In fact,  contrary to what happens with LQC and other non-covariant approaches such as teleparallelism or modified theories using the extrinsic curvature,
in our fully covariant approach the equations do not decouple, which is an added difficulty and ammounts to the fact that, in practice, the equations cannot be solved analytically but only by standard numerical methods.

\

A very positive feature is, however,  that for tensor perturbations our model leads to the same equations as GR because the modification of the action does not affect the gravity sector.

\

Finally, we  have studied the matter-ekpyrotic bouncing scenario for the LQC background,  when reheating is a consequence of the production of massless particles minimally coupled with gravity, during the phase transition from matter domination to the
ekpyrotic regime in the contracting phase. We have  obtained a  viable reheating temperature of the order of  $10^9$ GeV and have shown that
the observable modes leave the Hubble radius in the contracting phase, when the holonomy correction can be disregarded. This permits a very simple calculation of the  theoretical values of the spectral index  and of its running, both of which, as it turns out,  perfectly match the current observational data at the $2\sigma$ C.L..

\section*{Acknowledgments}
This investigation has been supported in part by MINECO (Spain), projects MTM2017-84214-C2-1-P and FIS2016-76363-P,  by the CPAN Consolider Ingenio 2010 Project, and by the Catalan Government 2017-SGR-247.

\end{document}